\documentclass[onecolumn]{aastex631}
\usepackage{threeparttable}

\newcommand{\msun}{$\rm{M_\odot}$ }
\newcommand{\msunns}{$\rm{M_\odot}$}
\newcommand{\bvf}{Brunt-V\"ais\"al\"a }
\newcommand\mh{$\mathrm{M_H}$ }
\newcommand\mhns{$\mathrm{M_H}$}
\newcommand\mhe{$\mathrm{M_{He}}$ }

\newcommand\menv{$\mathrm{M_{env}}$ }

\newcommand\teff{$\rm{T_{eff}}$ }

\shorttitle{WDEC versus LPCODE}
\shortauthors{Bischoff-Kim}

\begin{document}

\title{A comparison of best fits obtained in white dwarf asteroseismology using the WDEC and the LPCODE}

\correspondingauthor{Agn{\`e}s Bischoff-Kim}
\email{axk55@psu.edu}

\author[0000-0002-7487-9340]{Agn{\`e}s Bischoff-Kim}
\affil{Penn State Scranton \\
120 Ridge View Drive \\
Dunmore, PA 18412, USA}

\begin{abstract}

We perform the asteroseismic fitting of 4 DAVs using a grid of WDEC models with chemical profiles that closely mimic those of the LPCODE models and compare them with published asteroseismic fitting results using the LPCODE (Romero et al. 2017). These 4 objects are KIC 11911480, J113655.17+040952.6, KIC 4552982, and GD1212. The similarities in the results in those controlled experiments point to a consistency in the models. Given similar input, the LPCODE and the WDEC make similar models and calculate similar periods. We further perform the asteroseismic fitting of the same 4 DAVs by relaxing the constraints on the chemical profiles. We explore the effects of different methods for weighing the modes when calculating the goodness of fit of the models, as well as the effect of only including a subset of the known period spectrum. Such numerical experiments can help place recent and future efforts in pipeline fitting of numerous DAVs and DBVs using the WDEC on firmer footing \citep{Hall23}.

\end{abstract}

\section{Introduction} 
\label{sec:intro}

White dwarfs represent the final stage in the evolution of stars with masses less than 10 or 11 solar masses \citep[e.g.][]{Woosley15}. These compact remnants harbor fossil records within their interiors, providing insights into the vast majority of stars in our galaxy (approximately 98\%). The resulting chemical profiles play a crucial role in constraining physical processes such as nuclear fusion, core overshooting, mass loss, and diffusion.  During the core helium burning phase, nuclear reaction rates play a crucial role in determining the core composition of white dwarfs. Meanwhile, the relative duration of hydrogen and helium burning during the asymptotic giant branch (AGB) phase, along with mass-loss events, influences the thickness of the helium layer. The number of late thermal pulses (LTP's) determines the thickness of any hydrogen envelope \citep{Althaus02,Lawlor06,Althaus10,Miller-Bertolami16,Silva20}. The chemical profiles of the white dwarfs depend on the input physics. For instance, \citet{Metcalfe03b} propagated the errors on the NACRE rates \citep{Angulo99} for the $^{12}\mathrm{C(\alpha,\gamma)}^{16}\mathrm{O}$ reaction into the oxygen abundance profile of a 0.65 \msun white dwarf model. They found a range for the resulting central oxygen abundance between $\sim 0.55$ and  $\sim 0.75$. 

Most white dwarfs can be classified into two spectral types: DA (hydrogen-dominated atmospheres) and DB (helium-dominated atmospheres). Additionally, pulsating white dwarfs are designated with a “V” (e.g., DAV or DBV). Asteroseismology allows us to infer the interior structure of white dwarfs based on their light curves, while stellar evolution ties the structure to the underlying physics. Connecting the results of asteroseismology to the results of stellar evolution is an ongoing challenge. Only when the two are connected can we truly constrain the physics of stellar evolution with observational data. Efforts in the field fall on a spectrum, from codes that accept chemical profiles as input and then solve the stellar structure and energy transport equations to build a white dwarf model \citep[e.g.][]{Bradley98,Metcalfe02,Castanheira09,Fu13,Bognar16a,Giammichele18,Hall23}, to the use of stellar evolution codes to produce white dwarf models \citep[e.g.][]{Althaus08}. The former models are faster to run and allows flexibility in setting the chemical profiles. Because of this, they have been used to obtain models whose oscillations matched those of observed period spectra to an accuracy that reached the measurement uncertainties \citep{Giammichele18}.  The latter do not allow such a close match between model periods and observed periods, but they directly tie the white dwarf models to the physics involved in stellar evolution.

While a healthy scientific debate has taken place over the years \citep[e.g.][]{Fontaine02,Corsico19}, a true comparison of results obtained by different groups has not been published. There has been attempts to compare results from WDEC models \citep{Bischoff-Kim18a} and LPCODE models \citep{Althaus03}. One such comparison was performed in \citet{Althaus10}. In that work, the WDEC was used to produce chemical profiles that were similar to those of a chosen LPCODE fiducial model ($M_*=0.6096$ \msun, \teff $\sim$ 12 000~K), and a thick hydrogen envelope (\mh $\sim 10^{-4}$). At the time, limitations in the WDEC were such that we were unable to faithfully reproduce the LPCODE model. Yet, one noteworthy conclusion of the exercise was that the average period spacing of the WDEC model was smaller than that of the LPCODE model. The authors concluded that, should an asteroseismic fitting be performed on a grid of WDEC models that have chemical profiles more similar to those of LPCODE models, one would expect to find a best fit that is less massive and/or cooler than the equivalent LPCODE best fit. The two codes were pitted against each other again in analyzing the DBV TIC 257459955. Some agreement was found in the structure of the helium envelope between a good fit LPCODE model that was in agreement with the spectroscopy for the object and the second best fit WDEC model. The structure of the carbon and oxygen core in each case was completely different. The mass and effective temperatures of these best fit models was, respectively (0.609 \msun and 25 595~K) and (0.598 \msun, 24 546~K). 

With recent updates to the WDEC, we now have the ability to reproduce the chemical profiles of the LPCODE white dwarf models more faithfully (though not perfectly). In this work, we carry out the numerical experiment we wanted, but could not do before. We perform the asteroseismic fitting of 4 DAVs using a grid of models with chemical profiles that closely mimic those of the LPCODE models and compare them with published asteroseismic fitting results using the LPCODE \citep{Romero17}. These 4 objects are KIC 11911480, J113655.17+040952.6, KIC 4552982, and GD 1212. KIC 4552982 was the object of an asteroseismic fitting before with the WDEC \citep{Bell15}. \citet{Romero17} compared the results of the asteroseismic fitting of GD1212 with those obtained with WDEC models. We note, however, that the WDEC grid used in that work have cores made up of a homogeneous, 50/50 mix of carbon and oxygen \citep{Castanheira08}, very different from the profiles obtained from stellar evolution and from those used in this work. We will push the experiment further by performing the asteroseismic fitting of the same 4 DAVs by relaxing the constraints on the chemical profiles. Such numerical experiments can help place recent and future efforts in pipeline fitting of numerous DAVs and DBVs using the WDEC on firmer footing \citet{Hall23}.

We begin by a discussion of the models and of the numerical experimental procedure in section \ref{sec:methods}. We present the results in section \ref{sec:results}, and follow them with a discussion (section \ref{sec:discussion}). We summarize the work and highlight the important points in section \ref{sec:conclusion}.

\section{Methods}
\label{sec:methods}

\subsection{Models}
For this work, we use the WDEC, described in the instrument paper \citep{Bischoff-Kim18a}. Two improvements to the code have been made since that publication. One was an update in the equation of state tables. The code now interfaces with a newer version of MESA, version r22.11.1 \citep{Paxton22}. While making the code more state of the art and easier to install on modern architectures, it does not affect the white dwarf models to any significant level. Another change was to introduce a method in the parameterization of the chemical profiles that allows to have them default to a best reproduction of an LPCODE model, given a mass and effective temperature. The code defaults to the thickest hydrogen envelope model.

\begin{figure}[ht!]
\epsscale{0.6}
\plotone{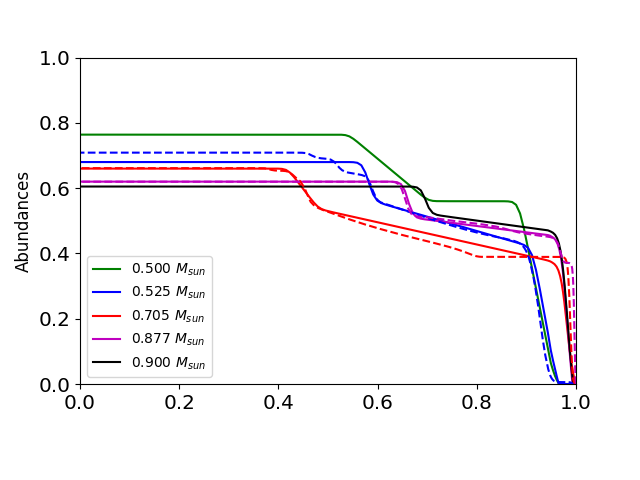}
\caption{Oxygen abundance profiles for LPCODE models (dashed lines) and WDEC models (solid lines) for different white dwarf masses. 
\label{fig:coprofiles}
}
\end{figure}

We show how well we are able to reproduce the oxygen profiles of the LPCODE models in Fig. \ref{fig:coprofiles}. The lowest mass LPCODE model available at the time the changes were made to the WDEC is 0.525 \msunns. We extrapolate our grids to masses as low as 0.500 \msunns. The higher mass LPCODE model is 0.877 \msunns, and we calculate models up to a mass of 0.900 \msunns. The four masses mentioned above are shown in Fig. \ref{fig:coprofiles}. We also show one intermediate mass, 0.705 \msunns.

\begin{figure}[ht!]
\epsscale{0.6}
\plotone{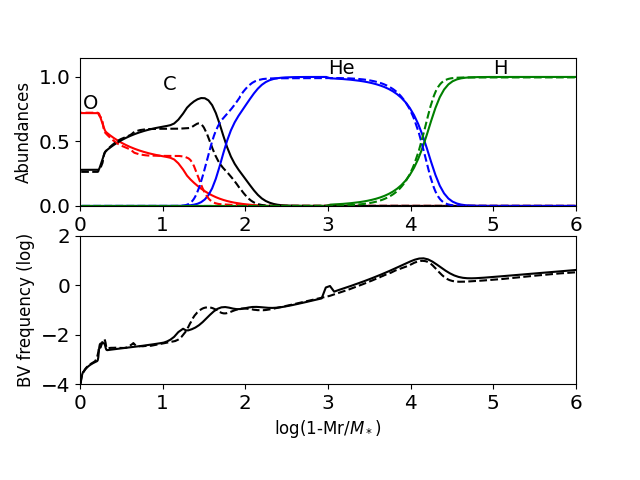}
\caption{Upper panel: composition profiles for a 0.605 \msun model, for the WDEC (solid lines) and the LPCODE (dashed lines). Lower panel: corresponding \bvf frequency curves. 
\label{fig:model_comparisons}
}
\end{figure}

In Fig. \ref{fig:model_comparisons}, we show the abundance profiles for oxygen, carbon, hydrogen, and helium for a 0.605 \msun model, again generated using the LPCODE and the WDEC. We changed the units on the horizontal axis to focus this time on structure in the helium and hydrogen envelope. In the lower panel, we graph the corresponding \bvf frequencies. The latter is what determines the period spectrum. The inner most bump in the \bvf frequency (the center is at 0), is a feature that has a strong influence on the period spectrum and we are able to reproduce that almost exactly with the WDEC. The bump at $\log(1-M_r/M_*)$ in the WDEC \bvf frequency is a numerical artifact. Upon close inspection, one can see a small discontinuity at that location in the helium abundance profile. That small discontinuity produces the bump in the \bvf frequency. This happens over the course of a few shells. \citet{Bischoff-Kim07} demonstrated that such discontinuity had an effect on the period spectrum of the order of a tenth of a second on the fitness parameter. There remains a limitation on how deep the helium layer can be. While we can push it to a depth of \mhe $\sim$ -1.20 when running an individual model, convergence does not always occur and so when running grids, we set the minimum helium layer mass to -1.80. 

Here it is useful to note that when defining composition profiles in the WDEC (where they go in as input), one provides a set of parameters to build the oxygen and helium profiles. The hydrogen abundance is calculated so that it begins to rise from zero at the outer edge of the helium layer all the way up to 1. The shape of the transition from helium to hydrogen is calculated assuming diffusive equilibrium. It is not parameterized. The carbon profile is calculated by subtracting the oxygen and helium abundances from 1, in the absence of any other element in that region of the model. Differences observed in the shape of the oxygen profiles arise from the fact that the oxygen profile parameterization does not allow the full complexity that arise from the time dependent diffusion calculations performed in the LPCODE.

\subsection{Grids}

Using the WDEC, we built 3 grids of models on which to perform our asteroseismic fittings. The first grid emulates an LPCODE grid, in that the only free parameters are the mass, effective temperature, and hydrogen layer mass of the models. The other two are grids one would build when approaching an asteroseismic fitting using the WDEC. In its most recent parameterization, the WDEC allows 16 free parameters (mass, effective temperature, 6 helium/hydrogen envelope parameters, 7 oxygen profile parameters, and a convective efficiency parameter). The parameters are described in detail in \citet{Bischoff-Kim18a} and \citet{Bischoff-Kim20}. Realistically, one cannot build a grid of sufficient resolution while varying 16 parameters. \citet{Bischoff-Kim23} did a systematic study aimed at determining which parameters mattered the most given different types of observed period spectra. They classified pulsation spectra according to the presence or absence of $\ell=2$ modes and the presence or absence of longer periods. The DAVs considered in this work fall in two types. KIC 4552982 and GD 1212 are of type 1, with rich spectra containing both $\ell=1$ and $\ell=2$ modes, as well as periods ranging from short to above 800~s. The resonant cavity of these longer period modes reaches all the way out to the base of the convection zone. The other two objects, KIC 11911480 and J113655.17+040952.6, are type 3. Their pulsation spectra present both $\ell=1$ and $\ell=2$ modes, but do not contain any periods above 500 seconds. They are also known as blue edge DAVs \citep{Mukadam06}. We detail each of the three grids in table \ref{tab:grids}.

\begin{table}
  \begin{center}
  \caption{Parameters used in the fits for the grid of models. The values listed for \menv are the negative of the log of the layer mass (e.g. "2" for "$10^{-2}$"). Similarly for \mhns.
     \label{tab:grids}
     }
 {\scriptsize
  \begin{tabular}{llll}
  \hline 
	Parameter & Minimum & Maximum & Step size      \\
    \hline
	\multicolumn{4}{c}{LPCODE grid}                \\
    \teff [K]       & 10000     & 13000     & 50      \\
    Mass [\msunns]  & 0.500     & 0.900     & 0.005    \\
    \mh             & 3.60      & 10.00     & 0.10     \\ 
    \hline
	\multicolumn{4}{c}{Type 1 grid}                \\
    \teff [K]       & 10000     & 13000     & 200      \\
    Mass [\msunns]  & 0.500     & 0.900     & 0.050    \\
    \menv           & 1.80      & 3.00      & 0.40     \\
    \mh             & 5.00      & 8.00      & 0.50     \\
    h1              & 0.50      & 1.00      & 0.05     \\
    w1              & 0.40      & 0.70      & 0.10     \\
    \hline	 
    	\multicolumn{4}{c}{Type 3 grid}            \\
    \teff [K]       & 10000     & 13000     & 200      \\
    Mass [\msunns]  & 0.500     & 0.900     & 0.050    \\
    \menv           & 1.80      & 3.00      & 0.40     \\
    \mh             & 5.00      & 8.00      & 0.50     \\
    h1              & 0.50      & 1.00      & 0.10     \\
    h2              & 0.30      & 1.00      & 0.10     \\
    w1              & 0.20      & 0.80      & 0.10     \\
    \hline
 \end{tabular}
 }
 \end{center}
\vspace{1mm}
\end{table}

\subsection{Fitting}
\label{sec:fitting}

For each model in the grid, we computed a goodness of fit statistic (Eq. \ref{eq:fiteq1}). At its core, this is a simple process of calculating a standard deviation between the observed list of periods and the corresponding periods from the model. The smaller the deviation, the better the fit. However, it is desirable to capture features of the observed pulsation spectrum, other than the period themselves. Some periods are measured with a higher precision than others. Some are more stable than others over time. Some modes have higher amplitudes than others. \citet{Romero17} chose to weigh the periods by their amplitudes. We see no good reason to do that, as amplitudes vary from observing run to observing run on a particular object, or from observing cycle to observing cycle in the extended coverage of space missions. We do not fully understand the variations in amplitudes, but the timescales (months) are such that we know that they are not due to a fundamental change in the internal structure of the white dwarf, which is what we are probing with asteroseismology. It is best to think of modes as being present or not present in the observed pulsation spectrum. If they rise above the chosen S/N threshold and are confirmed to be independent modes, they should count equally, regardless of their amplitude. 

It makes sense, however, to weigh modes according to how stable and precisely measured they are. Some modes, especially longer period modes \citep{Hermes17} present in the Fourier Transform not as sharp peaks, but rather envelops of power. \citet{Bell15} modelled these with Lorentzian curves and quoted for each mode, the location of the center of the Lorentzian bump, as well as its (half) width at half max (HWHM). This can serve as an error on the measurement of the period, which can then be folded into the calculation of the fitness statistic. We also have more traditionally determined errors on period measurements. Weighing periods with smaller errors (or HWHM) more heavily than those with higher errors de-emphasizes the importance of the modes that vary in period over time the most. Such modes are affected by the base of the convection zone and so less reliable than the more stable modes in informing the interior structure of the model. We adopt this latter weighing method. However, in the spirit of comparing with the results of \citet{Romero17}, we also compute fitness statistics weighing by amplitude. This allows us to gauge the effect of the different weighing methods on the results of the fitting procedures.

In summary, we calculate the goodness of fit statistic using the following formula:
\begin{eqnarray}
\label{eq:fiteq1}
\sigma_{\rm RMS} = \sqrt{\frac{1}{W} \sum_{1}^{n_{\rm obs}} {w_i(P^{\rm calc}_i-P^{\rm obs}_i)^2}}, \\
W=\frac{n_{\rm obs}-1}{n_{\rm obs}}\sum_{1}^{n_{\rm obs}}w_i
\end{eqnarray}

\noindent where $n_{\rm obs}$ is the number of periods present in the pulsation spectrum and the weights $w_i$ are either the inverse square of the errors on the periods, or the amplitude of each mode.

To further allow for comparisons that are as direct as possible, we adopt the mode identifications of \citet{Romero17}. This is not necessarily the mode identifications that lead to the best quality of fit with WDEC models. The purpose of this work is to see whether we recover similar stellar properties and internal structures if we hold as much as we can the same, rather than finding global best fits. When producing the contour plots shown in section \ref{sec:results}, we further took the inverse of the fitness parameter, such that good fits occur as maxima in those figures. This is again, to allow more direct comparison with the work of \citet{Romero17}. There is additional scaling that we are not reproducing, and so the comparisons are to be done in a relative sense. The absolute numbers are listed in tables \ref{tab:bestfits_kic11911480}-\ref{tab:bestfits_gd1212}.

\section{Results}
\label{sec:results}

\begin{figure}[h!]
\plotone{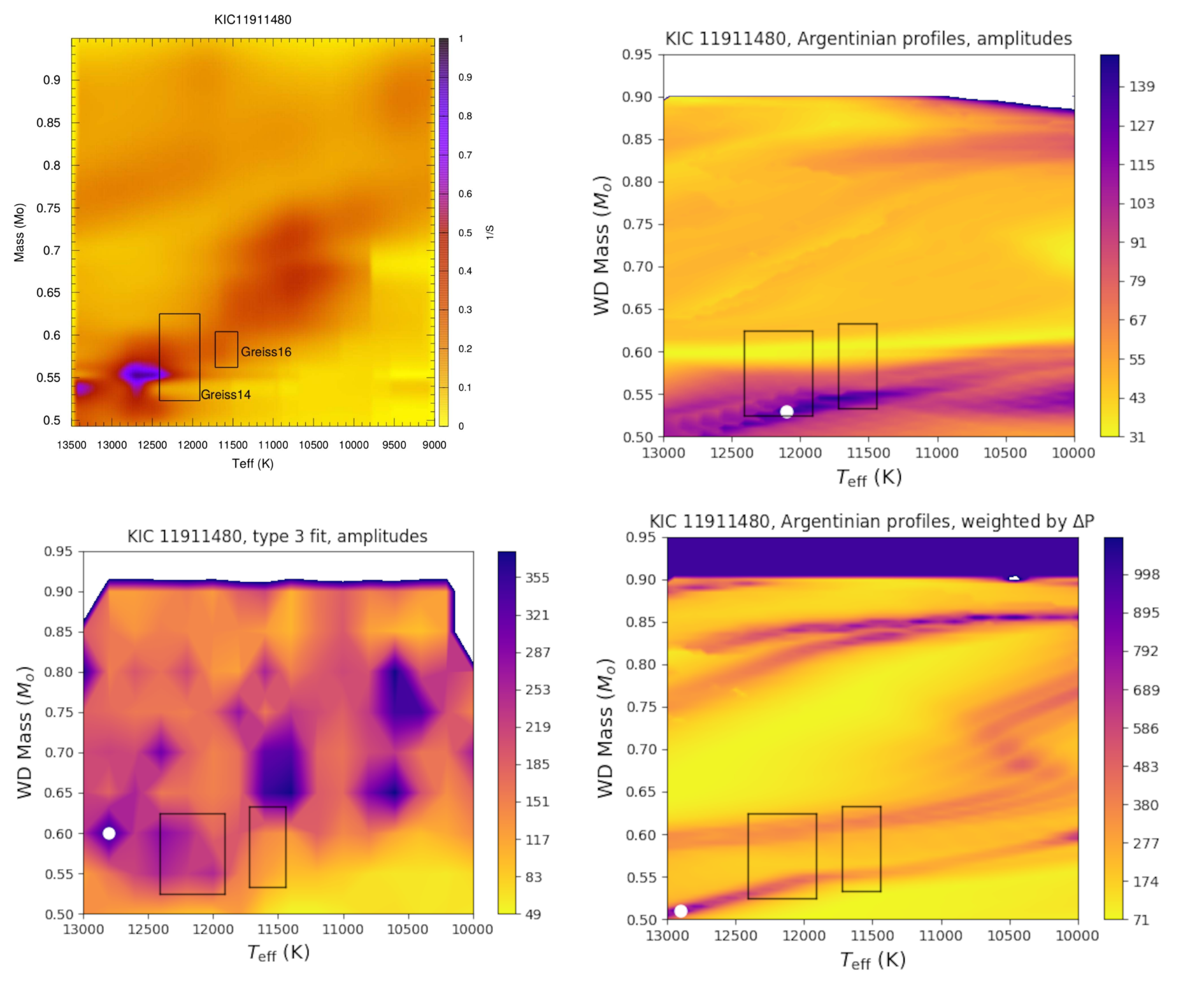}
\caption{Results of the fitting for KIC 11911480. Top left: Figure 2, reproduced from \citet{Romero17}. Top right: the same period spectrum, fitted on a grid of WDEC models that emulate the LPCODE grid, and weighting the fitness parameter by the amplitudes of the modes. Lower left: the same period spectrum fitted on the type 3 grid, with the fitness parameter weighed by amplitudes. Lower right: the same as the panel above it, except the modes are weighted by the errors on the periods. The spectroscopy is from \citet{Greiss14} and \citet{Greiss16}. We believe the Greiss16 spectroscopic box in \citet{Romero17} may have included the wrong error bars. \label{fig:kic11911480_massteff}}
\end{figure}

\begin{figure}[h!]
\plotone{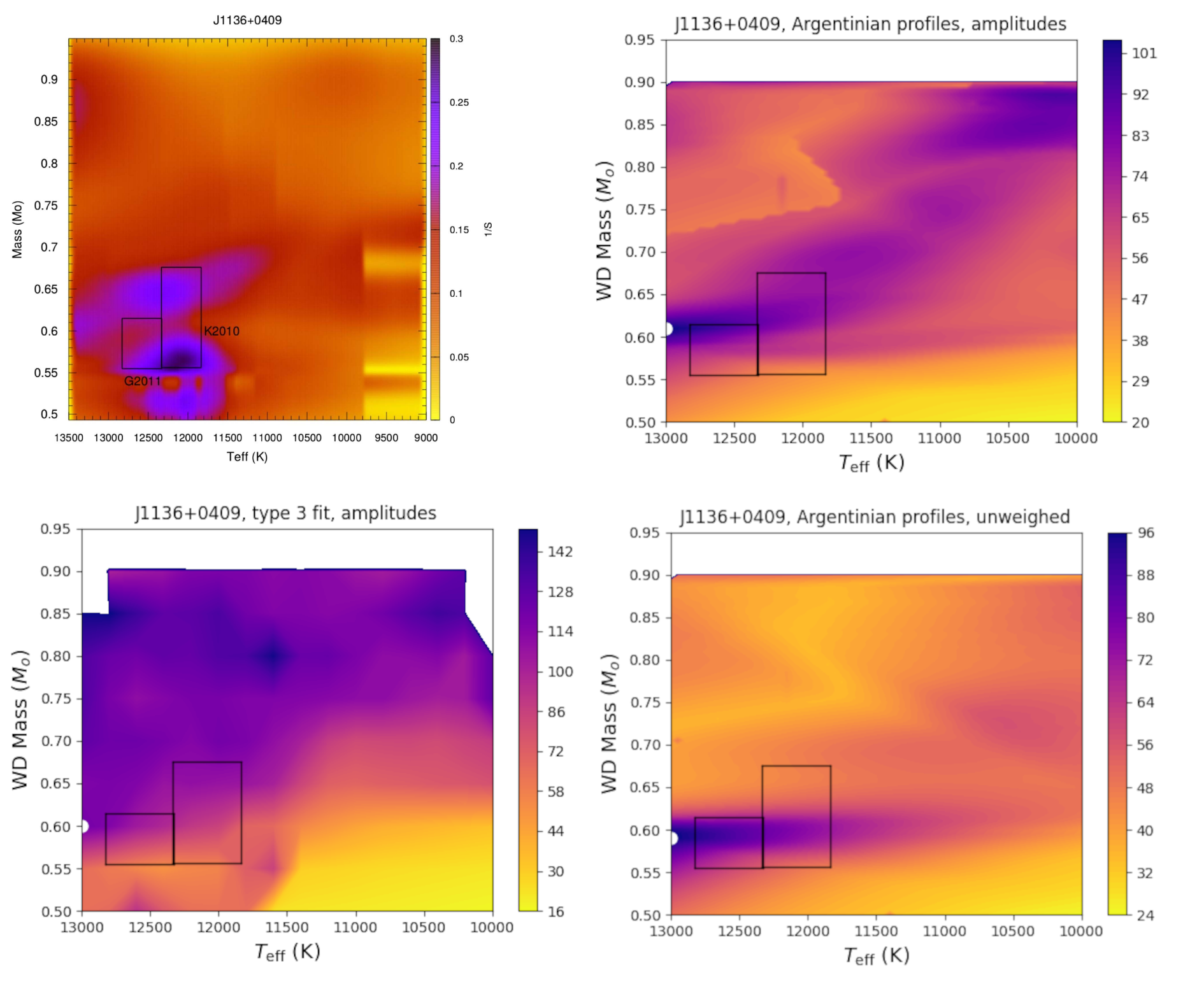}
\caption{Results of the fitting for J1136+0409. Top left: Figure 3, reproduced from \citet{Romero17}. Top right: the same period spectrum, fitted on a grid of WDEC models that emulate the LPCODE grid, and weighting the fitness parameter by the amplitudes of the modes. Lower left: the same period spectrum fitted on the type 3 grid, with the fitness parameter weighed by amplitudes. Lower right: the same as the panel above it, except the modes are all weighed equally. The spectroscopic boxes are from \citet{Gianninas11} and \citet{Koester10}. \label{fig:j1136_massteff}}
\end{figure}

\begin{figure}[h!]
\plotone{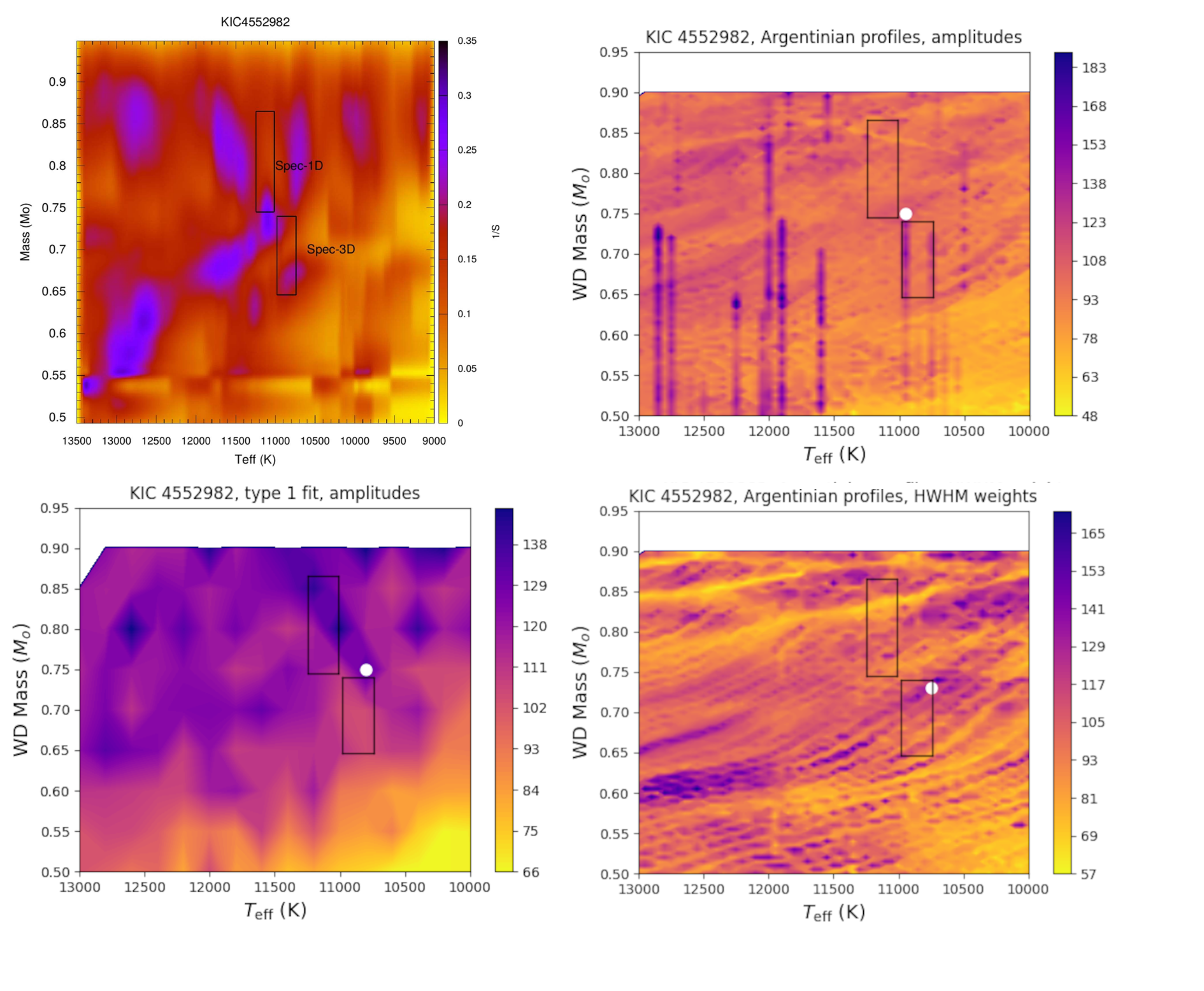}
\caption{Results of the fitting for KIC4552982. Top left: Figure 4, reproduced from \citet{Romero17}. Top right: the same period spectrum, fitted on a grid of WDEC models that emulate the LPCODE grid, and weighting the fitness parameter by the amplitudes of the modes. Lower left: the same period spectrum fitted on the type 3 grid, with the fitness parameter weighed by amplitudes. Lower right: the same as the panel above it, except the modes are weighted by the errors on the periods (defined as the half width at half max of the Lorentzian envelopes of power \citep{Bell15}). The spectroscopic boxes come from \citet{Hermes11} and \citet{Bell15} (SPEC-1D and SPEC-3D respectively). \label{fig:kic4552982_massteff}}
\end{figure}

\begin{figure}[h!]
\plotone{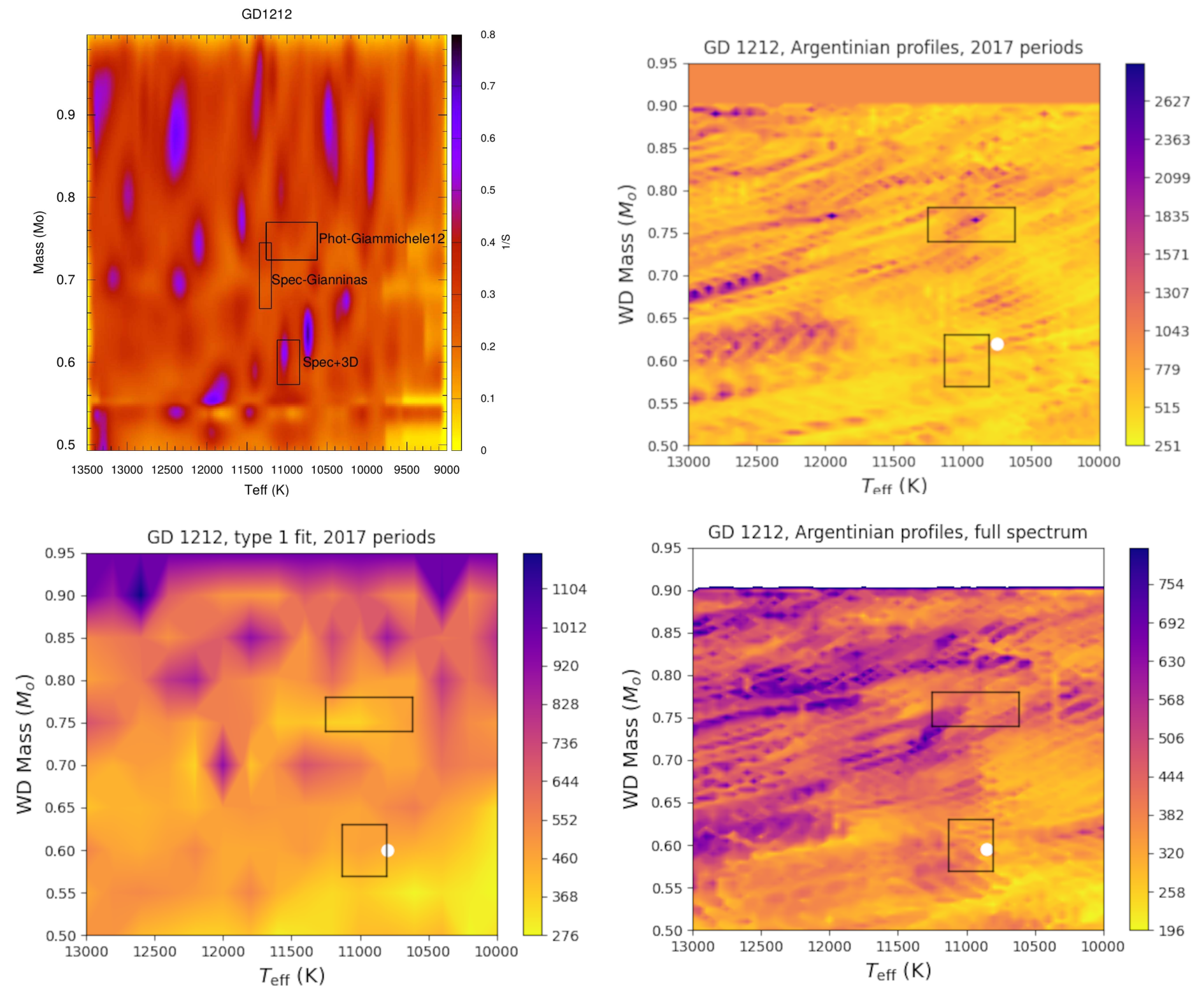}
\caption{Results of the fitting for GD1212. Top left: Figure 6, reproduced from \citet{Romero17}. Top right: the same period spectrum, fitted on a grid of WDEC models that emulate the LPCODE grid, and weighting the fitness parameter by the amplitudes of the modes. Lower left: the same period spectrum fitted on the type 1 grid, with the fitness parameter weighed by amplitudes. Lower right: the same as the panel above it, except fitting all known periods. The spectroscopic boxes come from \citet{Giammichele12} and \citet{Gianninas11}. \label{fig:gd1212_massteff}}
\end{figure}

We present below our results for each white dwarf in turn. For each, we performed a fit on the emulated LPCODE grid. In addition, depending on the type of pulsation spectrum, we performed a fit on a type 1 or type 3 grid (see table \ref{tab:grids}). We experimented with the weighing of the modes in calculating the goodness of fit. For GD 1212, we tried including all the known modes in addition to performing a fit of the 2017 subset used by \citet{Romero17}. In each contour map of Figures \ref{fig:kic11911480_massteff}-\ref{fig:gd1212_massteff}, we have marked with a solid white circle the location of a chosen best fit model and reproduced the spectroscopic boxes. With the exception of J1136+0409, the global minimum occurred at the highest mass (0.900 \msunns). The reason for this is that at higher masses, the period spacings becomes so small that there are more periods to choose from when trying to match the observed period spectrum. But none of the stars considered in this work are believed to have such high masses. As we note in section \ref{sec:discussion}, it is difficult to pinpoint unique solutions in the mass-effective temperature plane for KIC4552982 and GD1212. For these two objects, we selected strong local minima that occurred near the solutions found by \citet{Romero17} so that we could perform further comparisons.

\begin{figure}[h!]
 \epsscale{0.6}
\plotone{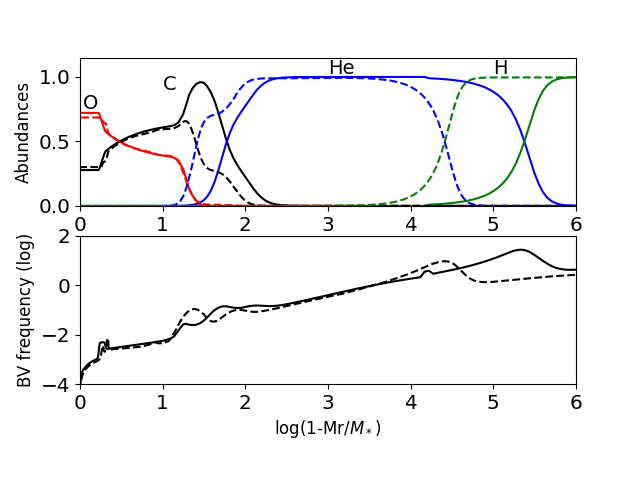}
\caption{Upper panel: composition profiles for the chosen best fit models for J1136+0409. The solid lines correspond to the WDEC model that results from performing the fitting on the LPCODE like grid, while the dashed lines correspond to the LPCODE model. Lower panel: corresponding \bvf frequency curves.. \label{fig:j1136_profiles}}
\end{figure}

In addition to producing best fit maps in the mass-effective temperature plane, we also compare the parameters of our best fit models. These are listed in tables \ref{tab:bestfits_kic11911480}-\ref{tab:bestfits_gd1212}. To illustrate how these parameters translate to composition and \bvf profiles, we plot the case of J1136+0409 in figure \ref{fig:j1136_profiles}.

\begin{table}
  \begin{center}
  \begin{threeparttable}[b]
  \caption{List of parameters characterizing the chosen best-fit models for KIC 11911480.
  \label{tab:bestfits_kic11911480}
}

  \begin{tabular}{lllll}
    \hline
    \hline
    Parameter           & LPCODE                & Argentinian profiles      & Type 3 grid               & Argentinian profiles      \\
                        &                       & Weighed by amplitudes     & Weighted by amplitudes    & Weighted by $\Delta P_i$  \\
    \hline  
    $M_*/M_\odot$       & 0.548                 & 0.530                     & 0.600                     & 0.510                     \\
    $T_{eff}$           & 12,721~K              & 12,100~K                  & 12,800~K                  & 12,900~K                  \\
    $\log g$            & 7.88                  & 7.89                      & 8.01                      & 7.85                      \\
    $\log(L/L_\odot)$   & -2.333                & -2.448                    & -2.417                    & -2.307                    \\
    $R/R_\odot$         & 0.0140                & 0.0136                    & 0.0126                    & 0.0141                    \\
    $log M_H$           & -3.419                & -7.400                    & -6.600                    & -6.100                    \\
    $log M_{He}$        & -1.117                & -1.800\tnote{*}           & -2.200                    & -1.800\tnote{*}           \\
    $X_c,X_o$           & 0.290, 0.697          & 0.292, 0.708              & 0.100, 0.900              & 0.283, 0.717              \\
    $S$                 & 1.13~s                & 7.00~s                    & 2.72~s                    & 0.993~s                   \\
   \hline
  \end{tabular}
    \begin{tablenotes}
    \item[*] Minimum allowed value in grid
  \end{tablenotes}
  \end{threeparttable}
 \end{center}
\vspace{1mm}
\end{table}

\begin{table}
  \begin{center}
    \begin{threeparttable}[b]
  \caption{List of parameters characterizing the chosen best-fit models for J1136+0409.
  \label{tab:bestfits_j1136}
}
  \begin{tabular}{lllll}
    \hline
    \hline
    Parameter           & LPCODE                & Argentinian profiles      & Type 3 grid               & Argentinian profiles      \\
                        &                       & Weighed by amplitudes     & Weighted by amplitudes    & Weighted equally          \\
    \hline  
    $M_*/M_\odot$       & 0.570                 & 0.610                     & 0.600                     & 0.590                     \\
    $T_{eff}$           & 12,060~K              & 13,000~K                  & 13,000~K                  & 13,000~K                  \\
    $\log g$            & 7.95                  & 8.02                      & 8.01                      & 7.98                      \\
    $\log(L/L_\odot)$   & -2.414                & -2.389                    & -2.388                    & -2.365                    \\
    $R/R_\odot$         & 0.0132                & 0.0126                    & 0.0126                    & 0.0130                    \\
    $log M_H$           & -4.507                & -5.163                    & -7.000                    & -4.800                    \\
    $log M_{He}$        & -1.212                & -1.800\tnote{*}           & -1.800\tnote{*}           & -1.800\tnote{*}           \\
    $X_c,X_o$           & 0.304, 0.696          & 0.278, 0.722              & 0.100, 0.900              & 0.302, 0.698              \\
    $S$                 & 2.83~s                & 9.56~s                    & 8.29~s                    & 10.4~s                    \\
   \hline
  \end{tabular}
      \begin{tablenotes}
    \item[*] Minimum allowed value in grid
  \end{tablenotes}
  \end{threeparttable}
 \end{center}
\vspace{1mm}
\end{table}

\begin{table}
  \begin{center}
  \caption{List of parameters characterizing the chosen best-fit models for KIC 4552982.
  \label{tab:bestfits_kic4552982}
}
  \begin{tabular}{lllll}
    \hline
    \hline
    Parameter           & LPCODE                & Argentinian profiles      & Type 1 grid               & Argentinian profiles      \\
                        &                       & Weighed by amplitudes     & Weighted by amplitudes    & Weighted by HWHM          \\
    \hline  
    $M_*/M_\odot$       & 0.745                 & 0.750                     & 0.750                     & 0.740                     \\
    $T_{eff}$           & 11,110~K              & 10,950~K                  & 10,800~K                  & 10,700~K                  \\
    $\log g$            & 8.26                  & 8.26                      & 8.27                      & 8.25                      \\
    $\log(L/L_\odot)$   & -2.815                & -2.841                    & -2.866                    & -2.873                    \\
    $R/R_\odot$         & 0.0105                & 0.0106                    & 0.0106                    & 0.0107                    \\
    $log M_H$           & -8.200                & -6.800                    & -7.000                    & -7.136                    \\
    $log M_{He}$        & -2.052                & -2.220                    & -2.200                    & -2.182                    \\
    $X_c,X_o$           & 0.343, 0.657          & 0.345, 0.655              & 0.150, 0.850              & 0.347, 0.653              \\
    $S$                 & 3.45~s                & 6.35~s                    & 8.92~s                    & 6.10~s                    \\
   \hline
  \end{tabular}
 \end{center}
\vspace{1mm}
\end{table}

\begin{table}
  \begin{center}
    \begin{threeparttable}[b]
  \caption{List of parameters characterizing the chosen best-fit models for GD1212.
  \label{tab:bestfits_gd1212}
}
  \begin{tabular}{lllll}
    \hline
    \hline
    Parameter           & LPCODE                & Argentinian profiles      & Type 1 grid               & Argentinian profiles      \\
                        &                       & Weighed by amplitudes     & Weighted by amplitudes    & All known periods         \\
    \hline  
    $M_*/M_\odot$       & 0.632                 & 0.620                     & 0.600                     & 0.595                     \\
    $T_{eff}$           & 10,737~K              & 10,750~K                  & 10800~K                   & 10,850~K                  \\
    $\log g$            & 8.05                  & 8.06                      & 8.03                      & 8.00                      \\
    $\log(L/L_\odot)$   & -2.737                & -2.753                    & -2.722                    & -2.688                    \\
    $R/R_\odot$         & 0.0123                & 0.0121                    & 0.0125                    & 0.0128                    \\
    $log M_H$           & -3.921                & -9.890                    & -6.000                    & -4.400                    \\
    $log M_{He}$        & -1.560                & -1.800\tnote{*}           & -3.000                    & -1.800\tnote{*}           \\
    $X_c,X_o$           & 0.245, 0.755          & 0.260, 0.740              & 0.000, 1.000              & 0.298, 0.702              \\
    $S$                 & 1.32~s                & 0.714~s                   & 2.26~s                    & 2.52~s                    \\
   \hline
  \end{tabular}
      \begin{tablenotes}
    \item[*] Minimum allowed value in grid
  \end{tablenotes}
  \end{threeparttable}
 \end{center}
\vspace{1mm}
\end{table}

\section{Discussion}
\label{sec:discussion}

We begin by contrasting the type 1 period spectra (GD1212 and KIC4552982) and the type 3 period spectra (J1136+0409 and KIC 11911480). The distinction between the two is the presence of long (greater than 800~s) periods in the type 1 pulsation spectra. As discussed in \citet{Bischoff-Kim23}, the presence of long period modes leads to multiple regions of best fits in the mass-effective temperature plane. This phenomenon is clearly seen here. For these stars, a unique determination of their mass and effective temperature from the periods alone is challenging and other constraints (for instance from spectroscopy) are required. For J1136+0409 and KIC 11911480, the constraints on mass and effective temperature are much stronger and for each object, we find a unique, global minimum (if one limits the grid to models below 0.800 \msunns, well above the expected mass for these two stars). It is these best fits that we list in tables \ref{tab:bestfits_j1136} and \ref{tab:bestfits_kic11911480}. In the introduction, we noted that we expected from earlier work \citep{Althaus10} that WDEC models with profiles similar to those of LPCODE models would point to best fits in the mass-effective temperature plane that were less massive and/or cooler. While  one may see a suggestion of that here, that question remains inconclusive. We were still unable in this work to faithfully reproduce the thickness of the helium envelope, and that has been shown to have an effect on the average period spacings \citep{Bischoff-Kim23}. 

One objective of this work is to test, for the first time, whether we recover the same solutions when using different codes. The results of this test are in the top two panels in Figures \ref{fig:kic11911480_massteff} - \ref{fig:gd1212_massteff}, and the answer is yes, qualitatively speaking. We do recover similar maps when holding as much as we can equal, aside from the code used to calculate the models and the period spectra. For GD1212 and KIC 4552989, that is harder to see, because the fits are ill constrained in the mass-effective temperature plane to begin with. However, the contour maps share qualitative features. It should be noted that the WDEC grid is much finer than the LPCODE grid. With only 3 parameters to vary and models that compute in seconds, we can afford that level of grid resolution. 

Aside for the codes used to compute the models and the pulsation periods, we also looked at the effect of differences in other aspects of the asterosesimic fitting: the weighing of each mode in the computation of the fitness parameter (Eq. \ref{eq:fiteq1}), and the effect of not including all observed periods. We tested the former with our fitting of KIC 11911480, J1136+-409, and KIC 4552982 and the latter with GD1212. \citet{Bischoff-Kim23} showed that choosing a subset of periods instead of the entire spectrum could lead to fits that were equal in quality, but did not point to the same internal composition profiles. Here we tried the experiment in a situation where the oxygen profile is fixed (as it is when using the stellar evolution approach to fitting) and find that the only parameter allowed to vary (the hydrogen layer mass) agrees between the two fits. We also constrained the mass and effective temperature to match, for reasons explained above. This is one object and a contrived situation and we still strongly advocate for using all known detected independent modes when doing asteroseismic fitting. Even in a 3 parameter fitting procedure (effective temperature, stellar mass, and hydrogen layer mass), there is a strong likelihood of not landing on the same solution. Consider the top right and bottom right panels of Fig. \ref{fig:gd1212_massteff}. There is a region of good fit that appears around (11,200~K,0.700 \msunns) that is not present when fitting only the 2017 period spectrum. Other features in the mass-effective temperature plane are similar.

We then push the numerical experiment further and consider a fit on a typical WDEC model grid, one where we vary the oxygen profiles, as well as the location of various transitions in the helium and hydrogen envelope (type 1 and type 3 fits). We still constrain the mass and effective temperature and mode identifications to match that found by \citet{Romero17}. For most stars, we recover helium and hydrogen layer masses, as well as central oxygen abundances that are similar to that of \citet{Romero17} (e.g. thin versus thick). The most consistent results are those for KIC 4552982, while the least consistent are for GD1212.

While in this work, we focused on comparing best fit models that had similar mass and effective temperature instead of global minima, we note the particularly bad quality of fits for J1136+0409. When we fitted that object with the WDEC \citep{Hermes15}, we found mode identifications for the best fit model that differed from that of \citep{Romero17} and the best fit model we chose then is not the (poorly fitting) local minimum listed in the 4th column of table \ref{tab:bestfits_j1136}.

\section{Summary and conclusion}
\label{sec:conclusion}

In this paper, we performed the asteroseismic fitting of 4 DAV's analyzed by \citep{Romero17}. Our main objective was to check the WDEC against the LPCODE. To that effect, we performed comparisons, trying to hold as much as we could in common. Throughout this work, we adopted the $\ell$ identifications of \citet{Romero17} and compared best fits found with WDEC to those found with the LPCODE that were close in the mass-effective temperature plane. The similarities in the results in those controlled experiments point to a consistency in the models. Given similar input, the LPCODE and the WDEC make similar models and calculate similar periods. This is the first time we have been able to check the two codes agains one another in this way. 

There is a lot of components to an asteroseismic fitting, beyond the codes used to run the models and calculate the oscillation periods. Parameterization and mode identification play a role as well. Taken together, these factors contribute to the discrepancies seen in the results of asteroseismic fittings from different groups. It is important to seek other constraints, such as spectroscopy and parallaxes \citep[e.g.][]{Bischoff-Kim24a}. Multiplet structure is valuable in aiding mode identification. Non-linear curve fitting {\citep{Montgomery10,Provencal12} is another method that can help narrow down the possibilities. One should not, however, succumb to the temptation of not including modes of unknown identification. 

The WDEC is open source and may be obtained from GitHub \citep[][Codebase: \\ \url{https://github.com/kim554/wdec}]{Bischoff-Kim18b}. Further documentation, including a user manual, may be found in the GitHub repository.

\software{WDEC \citep{Bischoff-Kim18b}}



\bibliography{index}{}
\bibliographystyle{aasjournal}

\end{document}